\documentclass[journal]{IEEEtran}

\usepackage{cite}
\usepackage{xcolor}

\ifCLASSINFOpdf
  \usepackage[pdftex]{graphicx}
  \graphicspath{{../pdf/}{../jpeg/}}
  \DeclareGraphicsExtensions{.pdf,.jpeg,.png}
\else
  \usepackage[dvips]{graphicx}
  \graphicspath{{../eps/}}
  \DeclareGraphicsExtensions{.eps}
\fi

\usepackage[cmex10]{amsmath}

\usepackage[caption=false,font=footnotesize]{subfig}

\usepackage{color}

\begin{document}
%
\title{A General Equivalent Circuit for Lossy Non-Symmetric Reciprocal Two-Ports}
%
%
%

\author{Alberto~Hern\'andez-Escobar,
        Elena~Abdo-S\'anchez,
        Pablo~Mateos-Ruiz,
        Jaime~Esteban,
        Teresa~M.~Mart\'in-Guerrero,
        and~Carlos~Camacho-Pe\~nalosa
\thanks{A.~Hern\'andez-Escobar, E.~Abdo-S\'anchez, Pablo Mateos-Ruiz T.~M.~Mart\'in-Guerrero and C.~Camacho-Pe\~nalosa are with the Departmento de Ingenier\'ia de Comunicaciones, Escuela T\'ecnica Superior de Ingenier\'ia de Telecomunicaci\'on, Universidad de M\'alaga, Andaluc\'ia Tech, 29010 M\'alaga, Spain (e-mail: ahe@ic.uma.es).}
\thanks{J. Esteban is with the Information Processing and Telecommunications Center, Escuela T\'ecnica Superior de Ingenieros de Telecomunicaci\'on, Universidad Polit\'ecnica de Madrid, 28040 Madrid, Spain.}
\thanks{This work was supported by the Spanish Ministerio de Educaci\'on, Cultura y Deporte (Programa para la Formaci\'on del Profesorado Universitario) under Grant FPU15/06457 and by the Spanish Ministerio de Ciencia, Innovaci\'on y Universidades (MCIU), the Agencia Estatal de Investigaci\'on (AEI) and the Fondo Europeo de Desarrollo Regional (FEDER) (Programa Estatal de
I+D+i Orientada a los Retos de la Sociedad) under grant RTI2018-097098-J-I00.}}

\markboth{}{}

\maketitle

\begin{abstract}
A general equivalent circuit for two-port lossy non-symmetric reciprocal networks is proposed. The equivalent circuit is based on the eigenstate decomposition. It has three complex parameters that are obtained from the eigenvalues and eigenvectors of the admittance or impedance matrix of the network in a straightforward way. The interconnection of the circuit elements is simple and compact. The applicability of the equivalent circuit is general, and the real parts of its immittances are always positive. To verify its behavior, the equivalent circuit of three different structures is extracted: an artificial composite right/left-handed transmission line unit cell, a series-fed coupled patch radiating element, and the complementary strip-slot. These cases offer a wide array of results, obtained from theoretical, simulated, and measured data. Nevertheless, the parameters of the equivalent circuit behave as expected in every case study in a very broad bandwidth. The equivalent circuit can be used advantageously in the analysis and design of non-symmetric two-ports, like unit cells of leaky-wave antennas.
\end{abstract}



\section{Introduction}\label{sec:intro}

\IEEEPARstart{T}{here} is a growing interest in the modeling of lossy non-symmetric reciprocal two-port electromagnetic structures by means of equivalent circuits. This interest has been fostered, to a great extent, by periodic Leaky-Wave Antennas (LWAs), since it has been shown that the introduction of certain asymmetries in their unit cell can mitigate the so-called open-stopband effect at broadside \cite{OTTO_2012, OTTO_2014}. The performance of the LWA is indeed governed by the asymmetries of the unit cell and their control can lead to improved LWA design strategies and behavior. Having a proper equivalent circuit can, in general, help provide some physical insight into the involved electromagnetic phenomena \cite{MESA_2018}, and can, in the particular case of LWAs, be crucial for extracting the propagation constant that determines the performance of such antennas \cite{MESA_2018_sym}.

From a purely mathematical point of view, three complex parameters are necessary for the complete specification of a dissipative reciprocal two-port structure, so the equivalent circuit must contain at least three lossy circuit elements or their equivalents \cite{Montgomery}. A given equivalent circuit is characterized by its constituent elements or components (mainly immittances, transformers, and transmission line sections or reference plane shifts) and its topology (the way the elements are interconnected). In the case of nondissipative two-ports there is a wide variety of available equivalent circuits (see, for instance, \cite{Montgomery,marcuvitz}). However, there is not such a variety of choices for the case of dissipative reciprocal two-ports.

One of the problems with dissipative reciprocal two-ports is that, depending on the topology of the equivalent circuit, immitances with negative real parts may appear, which is not a desirable characteristic, although it does not compromise energy conservation in the two-port as a whole. This modeling problem (the appearance of negative real parts in the immittances of the equivalent circuits) was already addressed in the early days of microwave technology. For instance, Felsen and Oliner \cite{FelsenOliner} further elaborated on an original idea from Weissfloch \cite{Weissfloch} and proposed to separate the dissipative equivalent circuit into lossy and lossless portions. In doing so, they obtained what they called \textit{canonical} networks or equivalent circuits composed by two transmission line sections (or two reference plane shifts), one reactance or susceptance, one ideal transformer with real turns ratio, and two non-negative resistors (a total of six elements, although only two of them are lossy). The idea of separating the dissipative two-port into lossy and lossless portions has reached our days and, very recently, Zappelli \cite{Zappelli} has proposed equivalents circuits for dissipative reciprocal two-port devices based on this separation approach. These equivalent circuits contain non-negative resistors and, in its simpler form (\textit{compact configuration}), has seven (real) elements and an intricate extraction procedure.

In the case of dissipative symmetric (certainly reciprocal) two-ports, the lattice network can be considered as the optimum equivalent circuit, since it guarantees the realizability of the two immittances required to model the response of any realizable symmetric network. Moreover, the lattice network provides a profound physical insight into the circuit behavior, since any response can be considered as a linear combination of the two \textit{eigenstates} (even- and odd-mode excitations) of the structure.  Because of this, in recent years, the spotlight has been put on the quest for lossy non-symmetric equivalent circuits that mimic, to some extent, the properties of the lattice network. Unfortunately, equivalent circuits based on the lattice-network topology require the structure to be symmetric, and the derivation of an extension of this topology for non-symmetric circuits has proved not to be straightforward.

In an attempt to elaborate on some design methodology to choose the degree of asymmetry needed to cancel the broadside effect in LWAs, an equivalent circuit based on the symmetric lattice network was proposed in \cite{OTTO_2014}. Although this equivalent circuit was useful for the purpose of \cite{OTTO_2014}, this circuit lacked the main aforementioned properties of the lattice networks which might limit its application to other problems. An alternative to this circuit is the equivalent circuit proposed in \cite{ELNv1}. The latter follows an eigenstate formulation approach \cite{Wane} and, unlike the circuit in \cite{OTTO_2014}, it preserves the property of symmetric lattice networks of being decomposable in eigenstates. This property of decomposition has been satisfactorily exploited very recently in \cite{LU_2019} to propose the design flow of a modified microstrip Franklin unit cell for its use in LWAs.

Although the benefits of the equivalent circuit proposed in \cite{ELNv1} have been already proved in specific LWA designs \cite{LU_2019}, this model has revealed to have two main problems. The first one is that the real parts of their immittances can be negative unless power orthogonality between the eigenstates is enforced by adding a frequency-dependent shift in the reference planes. This is an undesirable behavior for an equivalent circuit of a passive network. The second problem is that, when the two ports are isolated ($y_{12}$ or $z_{12}$ is zero or close to zero), the equivalent circuit is not able to properly degenerate into this limit case.

Based on the eigenstate decomposition proposed in \cite{ELNv1}, a novel equivalent circuit for non-symmetric structures is proposed here, which overcomes the main drawbacks of the equivalent circuits proposed there. The proposed circuit consists of two immittances and two transformers with a single complex turns ratio (i.e., three complex parameters, the minimum number of parameters required) as in \cite{ELNv1}, but with a different topology that has a dramatic effect on the immittances: their values are the eigenvalues of the immittance matrix of the structure. As an important consequence of this, the real parts of these two immittances are always positive (regardless of the real or complex character of the turns ratio or the choice of reference planes).  Additionally, the circuit is able to gracefully degenerate into the limit case defined by $y_{12}=0$ or $z_{12}=0$, which broadens its applicability and usefulness. Even more, the extraction procedure is absolutely straightforward, with simple explicit equations for all the circuit elements. Obviously, the equivalent circuit degenerates into the sought lattice network in the case of symmetric two-ports. The performance evaluation of the equivalent circuit is addressed by extracting its element values for three two-port cases: one based on the circuit simulation of an artificial CRLH unit cell, one based on the full-wave electromagnetic simulation of a series-fed coupled patch unit cell of a LWA, and one based on the measurements of a complementary strip-slot. The last two cases include an analysis of the influence on the equivalent circuit elements of the two-port degree of asymmetry.

This paper is structured as follows. Section II shows the eigenstate decomposition of a generic two-port. Section III proposes the new topology for both the admittance and impedance parameters. Section IV shows three examples of the application of the proposed equivalent circuit to model non-symmetric structures. Finally, Section V summarizes the main conclusions.


\section{Theoretical Background}\label{sec:background}

In \cite{ELNv1}, it was found that a non-symmetric reciprocal two-port network can be decomposed using an eigenstate formulation. For the sake of completeness, this argument is summarized here. Two decompositions were proposed, one of them using the admittance parameters and the other using the impedance parameters. From the admittance matrix of the network (the reasoning for the impedance matrix is analogous),
\begin{equation}\label{eq:1}
[Y]=\begin{bmatrix}
        y_{11} & y_{12} \\
        y_{12} & y_{22} \\
      \end{bmatrix}
\end{equation}
its eigenvalues can be obtained as
\begin{subequations}
\begin{equation}
\lambda_1=\frac{y_{11}+y_{22}+\sqrt{(y_{11}-y_{22})^2+4y_{12}^2}}{2}
\end{equation}
\begin{equation}
\lambda_2=\frac{y_{11}+y_{22}-\sqrt{(y_{11}-y_{22})^2+4y_{12}^2}}{2}
\end{equation}
\end{subequations}
and their associated normalized eigenvectors can be obtained as
\begin{subequations}
\begin{equation}
\overrightarrow{v_1}=\frac{1}{\sqrt{|p|^2+1}}\begin{bmatrix}
        p \\
        1 \\
      \end{bmatrix}
\end{equation}
\begin{equation}
\overrightarrow{v_2}=\frac{1}{\sqrt{|p|^2+1}}\begin{bmatrix}
        -1 \\
        p \\
      \end{bmatrix},
\end{equation}
\end{subequations}
where
\begin{equation}
p=\frac{y_{11}-y_{22}+\sqrt{(y_{11}-y_{22})^2+4y_{12}^2}}{2y_{12}}.
\end{equation}
By the definition of the eigenvalues and eigenvectors, it is possible to write, then,
\begin{subequations}
\begin{equation}\label{eq:xa}
[Y]\overrightarrow{v_1}=\lambda_1\overrightarrow{v_1}
\end{equation}
\begin{equation}\label{eq:xb}
[Y]\overrightarrow{v_2}=\lambda_2\overrightarrow{v_2}
\end{equation}
\end{subequations}
and, by interpreting these expressions in terms of circuits, the eigenvector $\overrightarrow{v_i}$ can be seen as applied voltages in each of the ports that produce the current $\lambda_i\overrightarrow{v_i}$ ($i=1,2$). Unifying the expressions (\ref{eq:xa}) and (\ref{eq:xb}) and separating the result into the sum of two matrices, it can be written as
\begin{equation}
[Y][\overrightarrow{v_1}\;\;\overrightarrow{v_2}]=[\lambda_1\overrightarrow{v_1}\;\;\overrightarrow{0}]+[\overrightarrow{0}\;\;\lambda_2\overrightarrow{v_2}].
\end{equation}
Then, it is possible to split the $[Y]$ matrix into two other admittance matrices, $[Y^a]$ and $[Y^b]$, as
\begin{subequations}
\begin{equation}
[Y]=[\lambda_1\overrightarrow{v_1}\;\;\overrightarrow{0}][\overrightarrow{v_1}\;\;\overrightarrow{v_2}]^{-1}+[\overrightarrow{0}\;\;\lambda_2\overrightarrow{v_2}][\overrightarrow{v_1}\;\;\overrightarrow{v_2}]^{-1}
\end{equation}
\begin{equation}
[Y]=[Y^a]+[Y^b].
\end{equation}
\end{subequations}
Their values are given by
\begin{subequations}
\begin{equation}\label{eq:2a}
[Y^a]=\lambda_1\begin{bmatrix}
        \frac{p^2}{p^2+1} & \frac{p}{p^2+1} \\
        \frac{p}{p^2+1} & \frac{1}{p^2+1} \\
      \end{bmatrix}
\end{equation}
\begin{equation}\label{eq:2b}
[Y^b]=\lambda_2\begin{bmatrix}
        \frac{1}{p^2+1} & \frac{-p}{p^2+1} \\
        \frac{-p}{p^2+1} & \frac{p^2}{p^2+1} \\
      \end{bmatrix}.
\end{equation}
\end{subequations}

Analogously, for the impedance formulation, the separation into submatrices is the following:
\begin{subequations}
\begin{equation}
[Z]=[Z^a]+[Z^b]
\end{equation}
\begin{equation}\label{eq:5a}
[Z^a]=\frac{1}{\lambda_1}\begin{bmatrix}
        \frac{p^2}{p^2+1} & \frac{p}{p^2+1} \\
        \frac{p}{p^2+1} & \frac{1}{p^2+1} \\
      \end{bmatrix}
\end{equation}
\begin{equation}\label{eq:5b}
[Z^b]=\frac{1}{\lambda_2}\begin{bmatrix}
        \frac{1}{p^2+1} & \frac{-p}{p^2+1} \\
        \frac{-p}{p^2+1} & \frac{p^2}{p^2+1} \\
      \end{bmatrix}.
\end{equation}
\end{subequations}
Since the admittance matrix is the inverse matrix of the impedance matrix, by the properties of the eigenvalues, the impedance-matrix eigenvalues are the inverse of the admittance-matrix eigenvalues and their eigenvectors are the same. For this reason, in this derivation, the same eigenvalues of the immittance matrix, $\lambda_1$ and $\lambda_2$, and the same relation between the coefficient of their eigenvectors, $p$, have been used, regardless of whether the admittance or impedance matrix is considered.

Several circuit topologies for the subnetworks $[Y^a]$, $[Y^b]$, $[Z^a]$ and $[Z^b]$ were proposed in \cite{ELNv1}. However, they were the cause of its two main problems. The first one is that
a frequency-dependent shift in the reference planes must be added
to obtain positive real parts for the circuit immitance. The second problem is that when ports 1 and 2 are isolated from each other ($y_{12}$ or $z_{12}$ are close to zero), none of the topologies in \cite{ELNv1} is capable to reproduce simultaneously the input impedances at the two circuit ports.

\section{Proposed Equivalent Circuit}\label{sec:new}

\begin{figure}[t]
\centering
\includegraphics[width=\columnwidth]{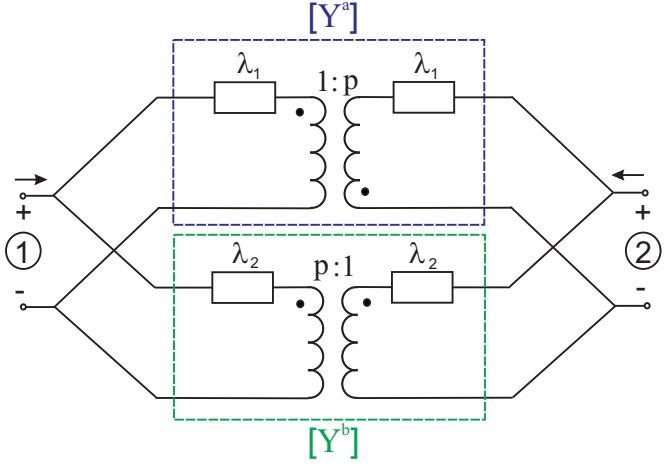}
\caption{Proposed eigenstate-based equivalent circuit using the admittance matrix.}
\label{fig:parallel}
\end{figure}

A new circuit topology for the subnetworks is proposed which solves the aforementioned drawbacks. The key point is to place a complex-turns-ratio transformer between to identical admittances, as depicted inside the dashed rectangles of Fig.~\ref{fig:parallel}, which has a dramatic impact on the performance of the equivalent circuit.

\begin{figure}[t]
\centering
\includegraphics[width=0.72\columnwidth]{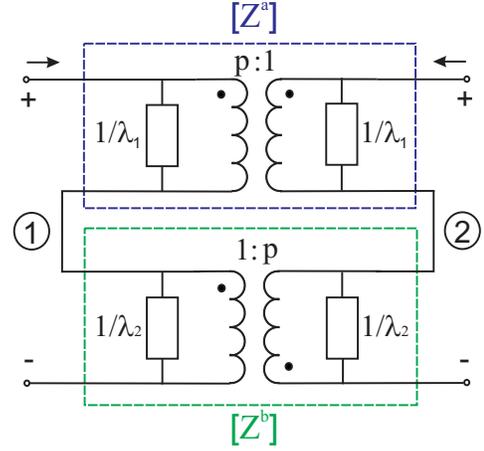}
\caption{Proposed eigenstate-based equivalent circuit using the impedance matrix.}
\label{fig:series}
\end{figure}

A suitable analogous topology for the impedance matrix case is also proposed and it is shown inside the dashed rectangles of Fig.~\ref{fig:series}. Note that, in both cases, the equivalent circuit is defined by three complex parameters, namely $\lambda_1$, $\lambda_2$, and $p$.

A relevant characteristic of these topologies is that the values of the immittances are the same as the eigenvalues of the admittance/impedance matrices. As proved in the Appendix, the real parts of the eigenvalues of the admittance matrix of a lossy reciprocal network are always positive or zero. Thus, every admittance of the equivalent circuit will never have negative real parts, provided the two-port is passive and reciprocal. Regarding the impedance-matrix-based equivalent circuit, the real parts of its impedances are always positive, since the inverse of a complex number with a positive real part will also have a positive real part.

Note that the turns ratios of the transformers are defined by a single parameter $p$, which is, in general, a complex number. It is important to note that complex-turns-ratio transformers are elements used in the synthesis of networks \cite{Huelsman}. It is also important to realize that these transformers are two-ports that, depending on the load conditions at their ports, can be lossy or active. If the turns ratio is real, the transformer is lossless. The role played by these two transformers in the equivalent circuit is clearly illustrated by computing the dissipated power in the two-port under arbitrary excitation. This power is given by:
\begin{multline}\label{eq:Power0}
P_{dis}=\frac{1}{2}Re\{V_1^*I_1+V_2^*I_2\}=\\
=\frac{1}{2}|A|^2Re\{\lambda_1\}+\frac{1}{2}|B|^2Re\{\lambda_2\}+\\
+\frac{1}{2}Re\Bigl\{\frac{(p-p^*)(A^*B\lambda_2-AB^*\lambda_1)}{|p^2|+1}\Bigr\},
\end{multline}which leads to:
\begin{multline}\label{eq:Power}
P_{dis}=
|A|^2P_{dis}^a+|B|^2P_{dis}^b+\\
+\frac{Im\{p\}}{|p^2|+1}Im\{AB^*\lambda_1-A^*B\lambda_2\},
\end{multline}where $A$ and $B$ are the complex amplitudes of the linear combination of eigenvectors that describe an arbitrary excitation, i.e.,
\begin{equation}\label{eq:AB}
    \begin{bmatrix}
        V_1 \\
        V_2 \\
    \end{bmatrix}=\frac{1}{\sqrt{|p^2|+1}}\Bigl\{A\begin{bmatrix}
        p \\
        1 \\
    \end{bmatrix}+B\begin{bmatrix}
        -1 \\
        p \\
    \end{bmatrix}\Bigr\}.
\end{equation}The first and second terms of (\ref{eq:Power}) are, respectively, the powers dissipated ($P_{dis}^a$ and $P_{dis}^b$) in each subcircuit when only the corresponding eigenmode is excited. It is important to highlight that $P_{dis}^a$ and $P_{dis}^b$ are always positive or zero as $Re\{\lambda_1\}$ and $Re\{\lambda_2\}$ are non-negative. It is straightforward to prove that, when only one of the eigenmodes is excited, the voltages at both ports of the corresponding transformer are zero (virtual short-circuit) and, thus, the dissipated power in the transformer is zero, even if the turns ratio is complex. The third term of (\ref{eq:Power}) is the so-called interaction term and accounts for the non-orthogonality of the eigenmodes. Note that this term is zero if the turns ratio is real, i.e., the eigenmodes are orthogonal. On the contrary, if the turns ratio is complex, the eigenmodes are non-orthogonal and the interaction term is non zero (positive or negative). In this case the complex-turns-ratio transformers are active or lossy and provide the required power to account for the energy conservation. The physical origin of the non-orthogonality of the eigenmodes is the asymmetry of the two-port. However, the asymmetry of the circuit does not necessarily imply the non-orthogonality of the eigenmodes. In particular, if the two-port is either resistive or lossless, the resulting turns ratios are real, which implies that the eigenmodes are orthogonal although the two-port is asymmetric.


Two limit cases should be mentioned, because of the capacity of the proposed circuit to properly represent both situations.
The first one is when $y_{12}=0$ ($z_{12}=0$) and the ports are isolated.
The turn ratios of the transformers tend either to zero or infinity ($p=0$ or $p\rightarrow\infty$) and therefore, the two transformers exhibit an open circuit and a short circuit in their ports, which isolates the two ports of both subnetworks.
As a result, the input admittance seen at each port will be $\lambda_1$ and $\lambda_2$, which are the theoretical values of the input admittances of the isolated two-port.
Given the value of $p$, the interaction term of (\ref{eq:Power}) vanishes, which corresponds to the fact that there is no interaction between one port and the other.

The second limit case is the symmetric network, with $p=1$.
The transformers then act either as a direct connection or as a crossover, and the resulting circuit topology becomes the classic lattice network with series admittances of value $\lambda_2$ and cross admittances of value $\lambda_1$.
Once again, the interaction term of (\ref{eq:Power}) vanishes (in this case because $p$ is real). An expected result, since the modes of the lattice network are always orthogonal.
Consequently, the lattice network becomes a particular case of the proposed equivalent circuit.


\section{Case Studies}\label{sec:cases}

In this section, the performance of the proposed equivalent circuit is evaluated by extracting its component values, over a wide frequency band, for three different electromagnetic structures.

\subsection{Artificial Composite Right/Left-Handed Transmission Line}

\begin{figure}[t]
\centering
\includegraphics[width=\columnwidth]{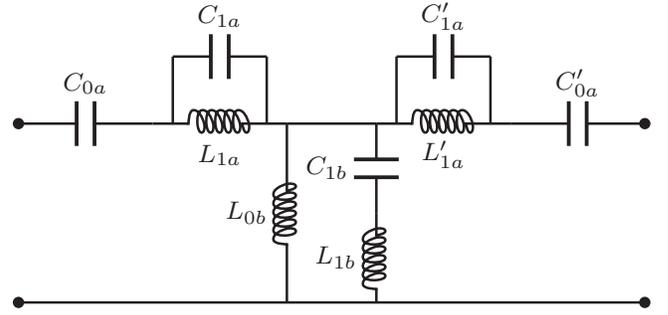}
\caption{Artificial CRLH transmission line unit cell. $C_{0a}=7.45$~pF, $L_{1a}=3.09$~nH, $C_{1a}=1.86$~pF, $C'_{0a}=5.59$~pF, $L'_{1a}=2.55$~nH, $C'_{1a}=2.05$~pF, $L_{0b}=15.38$~nH, $L_{1b}=6.21$~nH, $C_{1b}=1.23$~pF}
\label{fig:cellLDZ}
\end{figure}

\begin{figure*}[t]
\centering
\includegraphics[width=1.6\columnwidth]{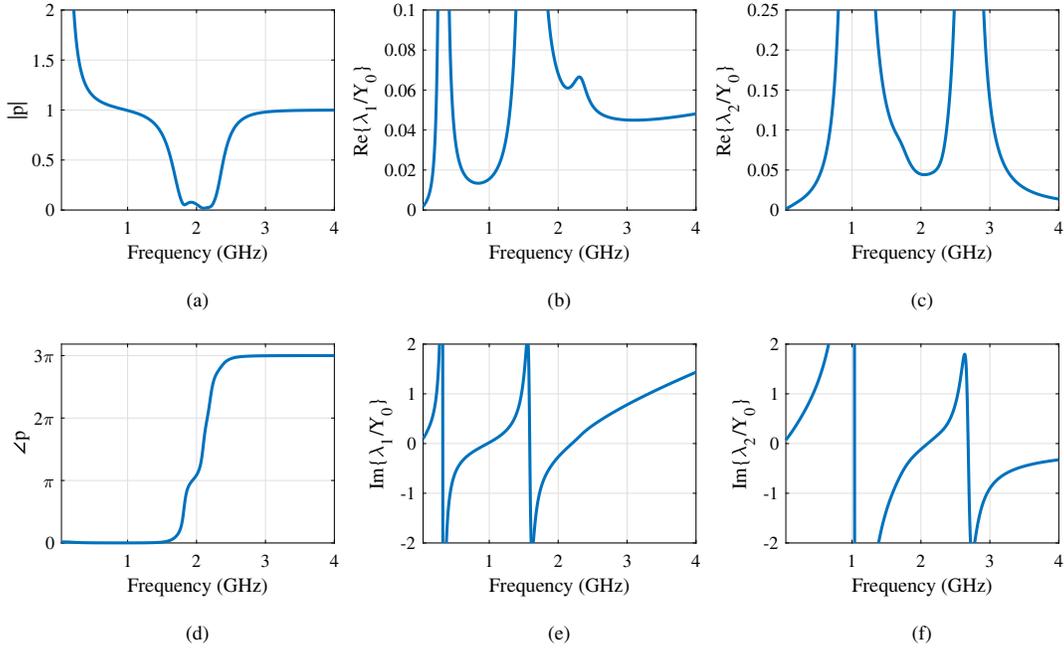}
\caption{Circuit parameters of the artificial CRLH transmission line unit cell using the proposed equivalent circuit. (a) Magnitude of $p$. (b) Real part of the normalized $\lambda_1$. (c) Real part of the normalized $\lambda_2$. (d) Phase of $p$. (e) Imaginary part of the normalized $\lambda_1$. (f) Imaginary part of the normalized $\lambda_2$. ($Z_0=1/Y_0=50~\mathrm{\Omega}$)}
\label{fig:resLDZ}
\end{figure*}

The first analyzed structure is the unit cell of an artificial composite right/left-handed (CRLH) transmission line. 
This first theoretical analytical case has been chosen to avoid any spurious phenomena coming from either simulation numerical noise or measurement uncertainties.
Specifically, the chosen unit cell is one of the two third-order artificial lossless CRLH transmission lines presented in \cite{ACRLH}: the \#6 unit cell in the taxonomy of \cite[Table 1]{ACRLH}. Like any other CRLH transmission line, it is said to be \textit{balanced} \cite{calozitho} when the distributed series impedance and the distributed shunt admittance have exactly the same critical frequencies (poles and zeros). The need for an exact match seems to be the main reason for the difficulties to achieve a perfect \textit{balance} in actual implementations and in avoiding the appearance of a stopband in the transition from left-handed to right-handed frequency bands.

To check the capability of the proposed circuit to analyze the asymmetries and imbalances of a real implementation of this unit cell, losses and asymmetries in the branches of the circuit have been considered. The losses have been included through a different finite quality factor for each element of the circuit (all quality factors around $Q\approx 50$). The asymmetries of the branches by means of slight modifications ($\approx \pm 10$\%) of the values of the elements of the circuit, which modify the resonance and anti-resonance frequencies of its immitances. The resulting circuit and element values are reproduced in Fig.~\ref{fig:cellLDZ}.

Fig.\ref{fig:resLDZ} shows the values of the components of the proposed equivalent circuit up to 4 GHz.  The analyzed structure has a stopband in the 1.8-2.2 GHz frequency band. At these frequencies, the $y_{12}$ parameter is almost negligible and, thus, $p$ is also close to zero. Even at this band, the admittances of the proposed equivalent circuit have positive real parts, and the change over frequency of the values of its circuit parameters is natural and well-behaved. Therefore, the capability of the proposed equivalent circuit to model structures that can have, over some bandwidths, decoupled ports is highlighted. Note that the transition between this stopband and the rest of the analyzed bandwidth is perfectly smooth. 

\subsection{Series-Fed Coupled Patch}

\begin{figure}[t]
\centering
\includegraphics[width=0.4\columnwidth]{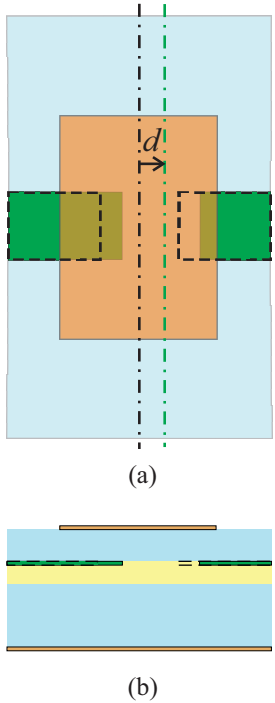}
\caption{The SFCP unit cell proposed in \cite{OTTO_2014}. (a) Top view. (b) Side view.}
\label{fig:cellSFCP}
\end{figure}

\begin{figure*}[t]
\centering
\includegraphics[width=1.6\columnwidth]{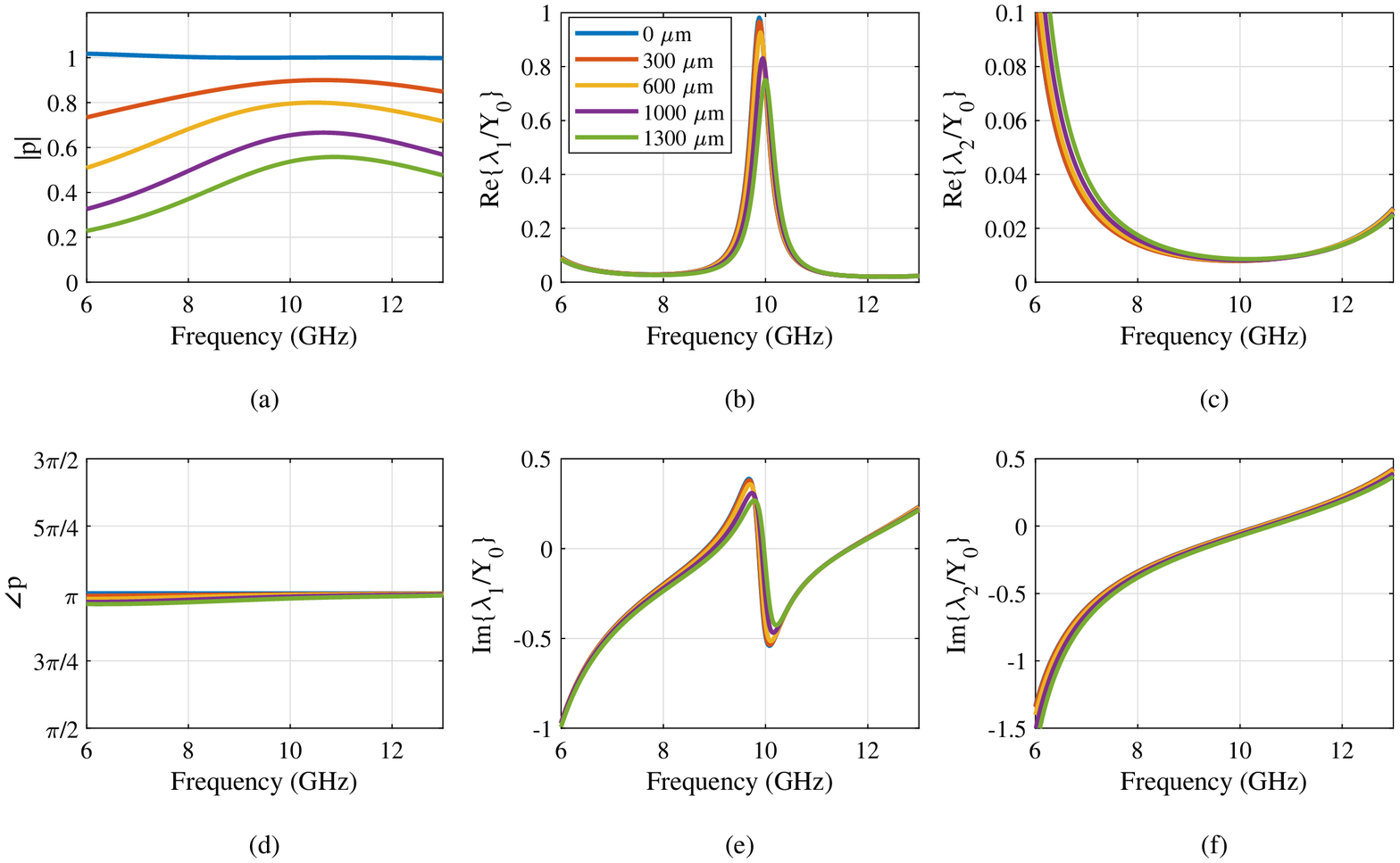}
\caption{Circuit parameters of the artificial SFCP unit cell for several asymmetries ($d$ parameter) using the proposed equivalent circuit. (a) Magnitude of $p$. (b) Real part of the normalized $\lambda_1$. (c) Real part of the normalized $\lambda_2$. (d) Phase of $p$. (e) Imaginary part of the normalized $\lambda_1$. (f) Imaginary part of the normalized $\lambda_2$. ($Z_0=1/Y_0=50~\mathrm{\Omega}$)}
\label{fig:resSFCP}
\end{figure*}

The second analyzed structure is the so-called Series-Fed Coupled Patch (SFCP) \cite{OTTO_2014}, which is studied through electromagnetic simulations using the commercial software HFSS. The SFCP consists of an on-top-stacked patch with gap coupling to the feeding line, as shown in Fig.~\ref{fig:cellSFCP}, and its dimensions are the same as in \cite{OTTO_2014}. In this case, the target of the study is how the parameters change when increasing its degree of asymmetry, which results from moving the patch along the feeding line axis a distance $d$. For this purpose, several unit cells with different degrees of asymmetry have been simulated and their equivalent circuit extracted.

Fig.~\ref{fig:resSFCP} shows the circuit parameters obtained when using the proposed equivalent circuit for several degrees of asymmetry. To have a reference for the degrees of asymmetry analyzed, if the value of $d$ were $d=1650\,\mu m$ the patch and microstrip would have no overlapping area in one of the sides, whereas $d=0$ corresponds to the symmetric case. 
In all cases, even with the numerical errors expected from a simulation, the real parts of the admittances are positive, and the equivalent circuit behaves smoothly also at lower frequencies, where the $y_{12}$ is low for this structure. Additionally, it can be seen that the values of the immittances barely change when the level of asymmetry is changed. The information about the asymmetry of the structure is mainly in the magnitude of the turns ratio of the transformers. This makes the new equivalent circuit a powerful tool to design LWAs using this unit cell.

\subsection{Complementary Strip-Slot}
In order to validate the previous results with actual measurements, the so-called \textit{complementary strip-slot} has been chosen as the last case study. 
This structure was proposed in \cite{ABDO_2012} as a planar radiating element with outstanding broadband matching. It consists of a slot etched on the ground plane of a microstrip line. 
The complementary stub (strip) of the slot is placed on the microstrip layer, symmetrically aligned to the slot, as a matching element. 
This structure is especially interesting for the present analysis because the contribution of the strip is associated with the even mode, and that of the slot, with the odd mode. 
A consequence of the eigenstate separation property of the proposed equivalent circuit is that it can separate the contribution of these two modes even if some asymmetry is added to the structure. This introduction of asymmetry to the structure was proposed in \cite{ABDO_2016} to reduce the broadside open-stopband effect when building an LWA with this element.

\begin{figure}[t]
\centering
\includegraphics[width=75mm]{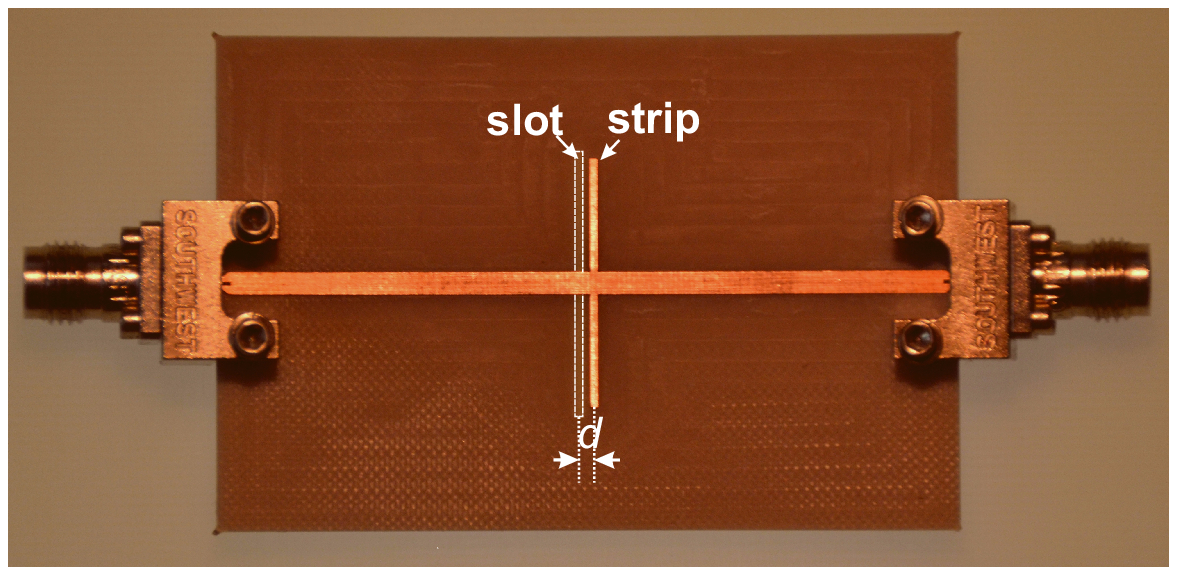}
\caption{Backlit photograph of a prototype of the complementary strip-slot with certain misalignment $d$.}
\label{fig:FotoPrototipo}
\end{figure}

\begin{figure*}[t]
\centering
\includegraphics[width=1.6\columnwidth]{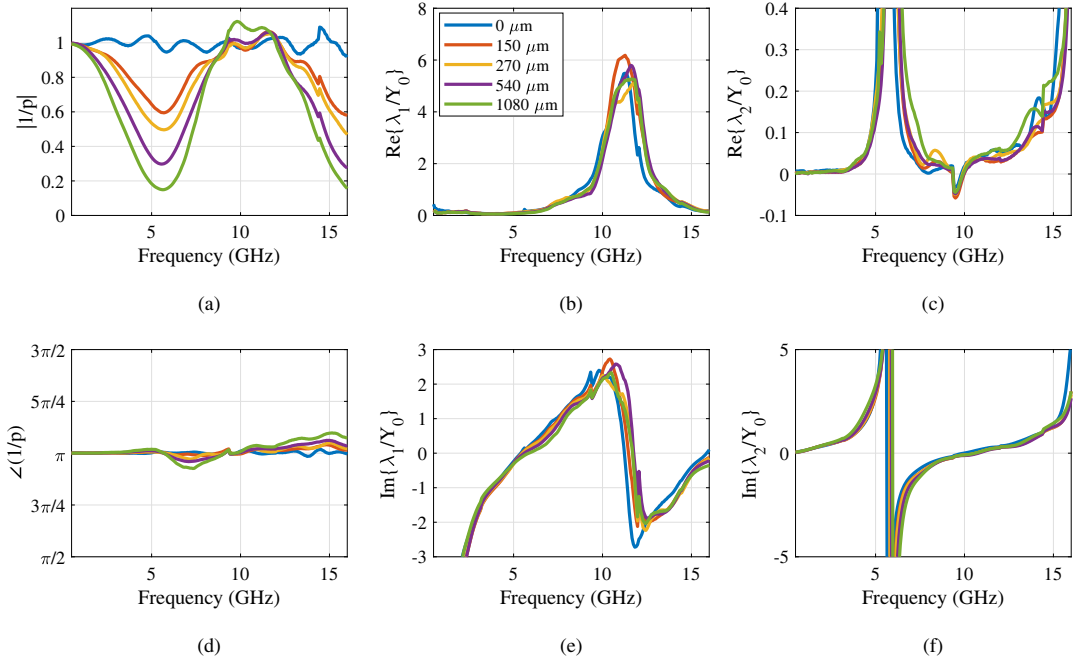}
\caption{Circuit parameters of the measured strip-slot unit cells for several asymmetries ($d$ parameter) using the proposed equivalent circuit. (a) Magnitude of $1/p$. (b) Real part of the normalized $\lambda_1$. (c) Real part of the normalized $\lambda_2$. (d) Phase of $1/p$. (e) Imaginary part of the normalized $\lambda_1$. (f) Imaginary part of the normalized $\lambda_2$. ($Z_0=1/Y_0=50~\mathrm{\Omega}$)}
\label{fig:resStripSlot}
\end{figure*}

Several prototypes of the strip-slot element with different degrees of misalignment $d$ have been fabricated. Fig. \ref{fig:FotoPrototipo} shows a partially backlit photograph of one of them. This misalignment is introduced by moving the strip and the slot a distance $d/2$ towards Port 2 and Port 1, respectively. In this case, the element values of the equivalent circuit have been extracted from the measured S-parameters with respect to $Z_0=50\,\Omega$, and the reference planes have been symmetrically placed $0.7$\,mm away from the center of the slot. To have a reference for the degrees of misalignment analyzed, the case of $d=540\,\mu m$ corresponds to the minimum value for which the strip and the slot have no overlapping area (they are adjacent elements loading the microstrip), whereas $d=0$ corresponds to the symmetric case. The lengths of the strip and the slot are such that their first resonant frequencies (total length equal to $\lambda/2$) appear around $5.4$\,GHz. The structure is fabricated on GML 1032 substrate with $\epsilon_r=3.2$ and $30$\,mil thickness.
The TRL calibration technique was used to obtain the measurements.
The calibration kit consisted of three “Lines” to cover the whole band of interest. Some anomalies occurred at the frequencies where the calibration changed from one “Line” to another, with the largest errors between 9.4 and 9.8 GHz.

Fig.~\ref{fig:resStripSlot} shows the equivalent circuit of the measured structures. In this case, the values of $p$ are greater than one and, in order to keep the shown results in the $0-1$ range, $1/p$ is plotted instead. 
Using the proposed circuit the real parts of the admittances are positive in all the frequency range, but between 9.4 GHz and 9.8 GHz. At these frequencies, the condition (\ref{eq:a2c}) from the Appendix is not fulfilled as a consequence of the anomalies introduced by the calibration kit. As in the previous case, the variation of the admittances with the asymmetry is small, taking into account that these are different physical structures. Thus, they still represent properly the slot and the strip admittances, separately, with the transformation ratio of the transformer absorbing the effect of the asymmetry. With this case study, the eigenstate-separation capabilities of the equivalent circuit are illustrated, in addition to its robustness in the presence of errors and inaccuracies from measurements.

\section{Conclusion}\label{sec:conclusion}

In this paper, a general eigenstate-based equivalent circuit for lossy non-symmetric reciprocal two-ports is proposed. It is general in the sense that it can be applied to any reciprocal lossy two-port regardless of its physical structure, since it does not rely on any physical aspect of the two-port. The circuit has only three complex parameters, namely two lossy immittances and the complex turns ratio of the transformers. It has been rigorously proved that the proposed circuit topology guarantees immittances with non-negative real parts. Moreover, the proposed equivalent circuit can efficiently handle two-ports with isolated ports ($y_{12}=0$ or $z_{12}=0$). Explicit and simple equations for the straightforward extraction of the equivalent circuit element values are provided.

The equivalent circuit has been validated by analyzing three different structures. In all cases, the results have proved and illustrated its capabilities and robustness against numerical and measurement uncertainties in a very broad band. An important feature of the proposed equivalent circuit is that it is capable of displaying the structure asymmetry explicitly. As a consequence, it is most convenient when the two-port has two identifiable eigenstates, since it easily models the underlying physics of the structure. This capability has been illustrated by two of the analyzed cases; in both, the turns ratio practically absorbs the asymmetry of the network leaving the immittances almost unperturbed. This is a desired feature when designing electromagnetic structures in which
slight asymmetries of the unit cell play an important role, as is, for instance, the case of the mitigation of the stopband in LWAs.

The excellent simplicity, eigenstate-based derivation and extraordinary behavior in modelling structures of different nature makes the proposed equivalent circuit a good candidate for modelling lossy non-symmetric reciprocal two-ports, especially when physical insight is pursued. Indeed, the highlighted capabilities of the proposed network suggest that it can be regarded as a generalization of the symmetric lattice network for lossy non-symmetric two-ports.  

\appendix\label{sec:appendix}

Let $[A]$ be a square matrix with complex elements and let $\pi\{[A]\}$, $\nu\{[A]\}$ and $\delta\{[A]\}$ be the number of eigenvalues of $[A]$ with positive real part, negative real part and zero real part, respectively. The triple pair $(\pi\{[A]\},\nu\{[A]\},\delta\{[A]\})$ is called the $inertia$ of $[A]$. According to \cite{OS62}, there exists a Hermitian matrix $[H]$ so that the matrix $[A][H]+[H][A]^*$ is positive definite if and only if $\delta=0$. In this case, $[A]$ and $[H]$ have the same $inertia$.

Let us apply this theorem to the admittance matrix of a passive and reciprocal circuit. First, it is possible to define the matrix formed by the real parts of the elements of $[Y]$ as
\begin{equation}\label{eq:a1}
Re\{[Y]\}=\frac{[Y]+[Y]^*}{2}=\begin{bmatrix}
        r_{11} & r_{12} \\
        r_{12} & r_{22} \\
      \end{bmatrix}.
\end{equation}
According to the energy theorem, the components of the admittance matrix must fulfill the following conditions \cite{Valkenburg}:
\begin{subequations}\label{eq:a2}
\begin{equation}
r_{11}\geq0
\end{equation}
\begin{equation}
r_{22}\geq0
\end{equation}
\begin{equation}\label{eq:a2c}
r_{11}r_{22}-r_{12}^2\geq0\,.
\end{equation}
\end{subequations}
Since the circuit is reciprocal, $Re\{[Y]\}$ is a symmetric matrix and, thus, its eigenvalues are real. Furthermore, if the circuit is lossy, $r_{11}>0$ and $r_{22}>0$, and then, by using the properties of the eigenvalues,
\begin{subequations}\label{eq:a3}
\begin{equation}
tr\Bigl\{\begin{bmatrix}
        r_{11} & r_{12} \\
        r_{12} & r_{22} \\
      \end{bmatrix}\Bigr\}=r_{11}+r_{22}=\lambda_{r1}+\lambda_{r2}>0
\end{equation}
\begin{equation}
det\Bigl\{\begin{bmatrix}
        r_{11} & r_{12} \\
        r_{12} & r_{22} \\
      \end{bmatrix}\Bigr\}=r_{11}r_{22}-r_{12}^2=\lambda_{r1}\lambda_{r2}>0,
\end{equation}
\end{subequations}
which implies that the $Re\{[Y]\}$ matrix is definite positive if the circuit is reciprocal and lossy, since its eigenvalues, $\lambda_{r1}$ and $\lambda_{r2}$, are real and positive.

It is possible to write (\ref{eq:a1}) as
\begin{equation}\label{eq:a4}
Re\{[Y]\}=\tfrac{1}{2}\left([Y][I]+[I][Y]^*\right)
\end{equation}
where $[I]$ denotes the identity matrix, which is Hermitian, and so, it is possible to apply the theorem from \cite{OS62}. This means that the $inertia$ of $[Y]$ and $[I]$ are the same. The eigenvalues of $[I]$ are $\lambda_{I1}=\lambda_{I2}=1$, real and positive, then,
\begin{equation}\label{eq:a5}
\pi\{[I]\}=2 \Rightarrow \pi\{[Y]\}=2.
\end{equation}
This way, it is proved that the real parts of the eigenvalues of the $[Y]$ matrix are always positive if the circuit is lossy and reciprocal.

If the circuit is lossless, it is straightforward to prove that the eigenvalues of $[Y]$ are purely imaginary numbers. The admittance parameters of a lossless circuit are always purely imaginary and, thus, it is possible to define a symmetric and real matrix $Im\{[Y]\}$ as
\begin{equation}
Im\{[Y]\}=\begin{bmatrix}
        i_{11} & i_{12} \\
        i_{12} & i_{22} \\
      \end{bmatrix}=-j[Y]
\end{equation}
where $i_{ij}$ are the imaginary parts of the circuit admittance parameters. Let us write the eigenstate decomposition as a function of this matrix:
\begin{equation}
j\begin{bmatrix}
        i_{11} & i_{12} \\
        i_{12} & i_{22} \\
      \end{bmatrix}\overrightarrow{v_i}=j\lambda_i\overrightarrow{v_i}
\end{equation}
where $\lambda_i$ is the eigenvalue of $Im\{[Y]\}$ and $\overrightarrow{v_i}$ the eigenvector of the $[Y]$ matrix. Since the $Im\{[Y]\}$ matrix is real and symmetric, its eigenvalues are real and, consequently, the eigenvalues of $[Y]$ are purely imaginary numbers.


\bibliographystyle{IEEEtran}
\bibliography{IEEEabrv,ELNv2.bib}

%
%








\end{document}